# Bulk Negative Index Photonic Metamaterials for Direct Laser Writing

Durdu Ö. Güney[1,*], Thomas Koschny[1,2], Maria Kafesaki[2], and Costas M. Soukoulis[1,2]

[1]Ames National Laboratory, USDOE and Department of Physics and Astronomy, Iowa State University, Ames, IA 50011, USA

[2] Institute of Electronic Structure and Laser, Foundation for Research and Technology Hellas (FORTH), and Department of Materials Science and Technology, University of Crete, 7110 Heraklion, Crete, Greece

*Corresponding author: dguney@ameslab.gov

**Abstract:** We show the designs of one- and two-dimensional photonic negative index metamaterials around telecom wavelengths. Designed bulk structures are inherently connected, which render their fabrication feasible by direct laser writing and chemical vapor deposition.
©2008 Optical Society of America

Simultaneously negative effective magnetic permeability and electric permittivity of metamaterials gives rise to exotic electromagnetic phenomena [1—4] not known to exist naturally and these materials enable a wide range of new applications as varied as cloaking devices and ultrahigh-resolution imaging systems. All photonic metamaterials at THz frequencies have been fabricated by well-established 2D fabrication technologies such as electron-beam lithography and evaporation of metal films, and most of them are only one or two functional layers [5—13]. A few efforts have been made to fabricate three to five layers [14—16], but this is also a 1D design. However, isotropic 3D bulk negative index metamaterial (NIM) designs with low absorption and high transmission that operate at THz and optical frequencies are needed to explore all the potential applications of NIMs. Direct laser writing (DLW) is a more promising technique for the fabrication of truly 3D large-scale photonic metamaterials. Rill *et al* recently demonstrated the feasibility of this technique at near infrared frequencies assisted with silver chemical vapor deposition [17]. In this letter we present the first truly bulk NIM design, which is feasible to fabricate with DLW at around telecom wavelengths.

Initial demonstrations of negative index metamaterials have been performed at GHz ranges by using composite structured materials consisting of planar split ring resonators (SRRs) and wires deposited on dielectric substrates [18]. Numerical studies have revealed that NIMs with reduced loss at near-infrared (NIR) regions can indeed be designed to be fabricated with standard optical interferometric lithography [19]. Subsequent efforts [5—9] led to the first experimental verifications of NIMs at optical frequencies despite a low figure of merit (FOM) [defined as $-\text{Re}(n)/\text{Im}(n)$ at $\text{Re}(n) = -1$]. Using a structure consisting of double-plate pairs and long wires, relatively higher FOM of about 3 was achieved [10,11] experimentally in 2006, which was later miniaturized [12,13] to operate in the visible spectrum (i.e., 780nm). Designs of 3D isotropic NIMs exist but fabricating them has remained to be a challenging task [20,21]. Only 2D isotropic NIMs are shown to be possible at microwave frequencies, where the propagation is either in two directions with a fixed polarization or in one direction with an arbitrary polarization [20,22,23]. However, these designs can not be miniaturized to give NIMs at THz and optical frequencies. Below we describe new designs of both one-dimensional (1D) and 2D isotropic negative index photonic metamaterials consisting of interconnected bulk SRRs. Inherent interconnections render the fabrication feasible by direct laser writing and chemical vapor deposition. Lithography of bulk polymeric templates by DLW has widely been used [24—26] with sizes of less than 100 nm. Having the template interconnected one can use chemical vapor deposition to cover it with a thin metallic film and therefore one will have a 3D photonic metamaterial.

The design for 1D photonic NIM is illustrated in Fig. 1. It consists of a pair of gold cylindrical SRRs (radius 37 nm) embedded in a polyimide substrate and separated by a distance of 120 nm in $z$-direction. The Drude model is used to describe the metal. The plasma and collision frequencies for bulk gold are assumed[15] ($\omega_p = 2\pi \times 2.175 \times 10^{15}$ s$^{-1}$, $\omega_c = 2\pi \times 6.5 \times 10^{12}$ s$^{-1}$). The dimensions of the unit cell are 220 nm $\times$ 240 nm $\times$ 320 nm. All the bent cylindrical parts have the same curvature of 2/75 nm$^{-1}$ at their center lines. The unit cell is designed to have inversion symmetry to allow for the homogeneous effective medium (HEM) approximation and retrieval of effective refractive index, electric permittivity, and magnetic permeability. See Fig. 1b for other parameters.

The incident field is configured (see Fig. 1a) in such a way that the electric field and the magnetic field couple to the SRR gap and the inner loop, respectively. We use perfect electric conductor (PEC) and perfect magnetic conductor (PMC) boundary conditions. Under these settings magnetic resonance originates from the combined capacitance and inductance of the nanocircuit. The former results from the individual SRR gaps and close proximity of the two wires (see Fig. 1b), while the inner loops provide the latter. The plasmonic response, on the other hand, arises from parallel currents oscillating along the continuous wires.

All the simulations were performed with the CST Microwave Studio software package (Computer Simulation Technology GmbH, Darmstadt, Germany). Vacuum wavelength to unit cell size in the propagation direction is more than 6 at 215 THz (i.e., can be approximated by a HEM). The retrieved effective parameters [27] are shown as a function of frequency in Fig. 1 (c-e). Notice that both $\varepsilon$ and $\mu$ are negative and therefore $n$ is also negative. The structure shown in $\varepsilon$ and $\mu$ and the imaginary part of $\varepsilon$ being negative is due to periodic effects [28]. The designed metamaterial structure has a negative index region of a bandwidth of 30 THz around the optical communication wavelengths. Numerically calculated figure of merit is 10 at 215 THz where $n = -1$ (not optimized).

In literature, it is the single unit cell retrieval which is usually shown for metamaterials. Although this is necessary to check in the first place whether the proposed structures manifest the desired properties, it is not sufficient for a realistic large-scale metamaterial, because the corresponding multi-cell structure may not necessarily reproduce the similar electromagnetic properties expected from a single unit cell.

In Fig. 2 we display the multiple cell results for the above 1D metamaterial and confirm that the structure still preserves its negative index property in the region of interest (around 215 THz). In Fig. 2 the retrieved index of refraction up to 11 unit cells are plotted in the same graph. Above 200 THz the structure can be reasonably well approximated by a HEM and thus the results from the retrieval procedure for multiple unit cells closely follow the single unit cell retrieval results and tend to coincide except for the initial shift from single unit cell to two unit cell curves. At the low frequency end of Fig. 2, where index of refraction is positive, individual curves again tend to coincide (not shown). However, at intermediate frequencies due to the periodicity artifacts and the resonance the effective medium approximation starts to break down and leads to relatively erratic behavior. Therefore we truncated the curves for multiple unit cells in this region. Also the retrieved effective electric permittivity and magnetic permeability (not shown) for the last four (i.e., 8-11) multiple unit cells give converging results. It would be desirable for both HEM representation and practical applications that the negative index region is clear from any periodicity artifacts, which could be achieved by shrinking the unit cell while maintaining the same resonance frequency. Nevertheless, we should remind that the designed structure has still room for additional optimization and therefore it may be possible to push these artifacts away from the region of interest.

Having discussed the 1D structure, we show that using essentially the same SRRs with a slightly different geometric configuration it is possible to obtain a negative index of refraction in two propagation directions. For this purpose, designed 2D NIM unit cell has C4 symmetry as illustrated in Fig. 3. It consists of one pair of SRRs with inversion symmetries in both *x*- and *z*-directions similar to 1D unit cell in Fig. 1. The structure has a square lattice of lattice constant $a = 364$ nm. All the other parameters are chosen to be the same as 1D case. Incident field is configured as shown in Fig. 3a. We apply PEC and PMC boundary conditions to impose periodicity in non-incident directions. Although two-dimensionally periodic structures cannot necessarily be simulated by applying PEC/PMC boundary conditions, it turns out that the electromagnetic field cannot actually distinguish the symmetry in this specific structure. We checked this by applying periodic boundary conditions and observed that the change is negligible. Compared with periodic boundaries, the PEC/PMC boundaries have advantage that they are generally more efficient in terms of time and memory with the CST Microwave Studio.

The vacuum wavelength to unit cell size in propagation directions is about 5. The retrieved effective refractive index, magnetic permeability, and electric permittivity using HEM approximation are displayed in Fig. 3 (c-e). Although the structure has room for further optimization, it already has a figure of merit of about 5 and the operation frequency lies close to telecom band (150 THz) with about 20 THz bandwidth. The designed structure has a 2D isotropy in the sense that the propagation is in two directions with fixed polarization and index of refraction.

In Fig. 4 we plot the multiple cell simulation results up to four unit cells. In contrast to 1D structure, the 2D one already homogenizes after a few number of unit cells. This may be the result of relatively large separation of metal cores in the 2D square lattice compared to the 1D lattice which may have weakened the edge effects between neighboring unit cells.

Like the 1D structure, however, HEM approximation again starts to break down around the resonance below 160 THz and recovers itself at lower frequencies. Fig. 4 shows the effective refractive index for different lengths of unit cells, and one sees clearly how good the curves coincide. Also the corresponding retrieved effective electric permittivity and magnetic permeability are negative and well coincide too. No periodicity artifacts are observed in the frequency band of interest in the 2D case as opposed to the 1D.

We present the designs of both 1D and 2D bulk photonic metamaterials which exhibit negative index of refraction around the optical communication wavelengths. Given the state-of-art 2D fabrication technologies manufacturability is still a concern to realize any practical optical devices based on metamaterials. Therefore alternative technologies such as direct laser writing with compatible designs are essential. Our design due to the inherently connected nature of the SRRs with their next-unit-cell neighbors presents the first photonic metamaterials feasible to fabricate with direct laser writing and chemical vapor deposition. This may pave the way to fabricate large-scale 3D metamaterials operating at optical frequencies and therefore all the potential applications can become reality.

Work at Ames Laboratory was supported by the Department of Energy (Basic Energy Sciences) under contract No. DE-AC02-07CH11358. This work was partially supported by the AFOSR under MURI grant (FA9550-06-1-

**Figure 1** (Color online) (**a-b**) Unit cell with parameters and the incident field configuration for the 1D NIM consisting of one pair of SRRs and (**c-e**) the corresponding retrieved effective parameters using HEM approximation. Solid red lines indicate the real parts and dashed blue lines indicate the imaginary parts. Magnetic permeability and electric permittivity are simultaneously negative in the region of negative index of refraction.

**Figure 2** (Color online) Retrieved effective index of refraction for different lengths of unit cells (1 to 11) plotted in the same graph. Curves for multiple unit cells (thin lines) tend to coincide except for the initial shift from the single unit cell curve (thick blue line) to the two unit cell curve.

**Figure 3** (Color online) (**a-b**) Unit cell and the incident field configuration for the 2D NIM consisting of two pairs of SRRs and (**c-e**) the corresponding retrieved effective parameters using HEM approximation. Solid red lines indicate the real parts and dashed blue lines indicate the imaginary parts. Magnetic permeability and electric permittivity are simultaneously negative in the region of negative index of refraction.

**Figure 4** (Color online) Retrieved effective index of refraction for the first four lengths (i.e., 1 to 4 unit cells) plotted in the same graph. Curves for multiple unit cells (thin lines) coincide even with that of single unit cell (thick blue line).

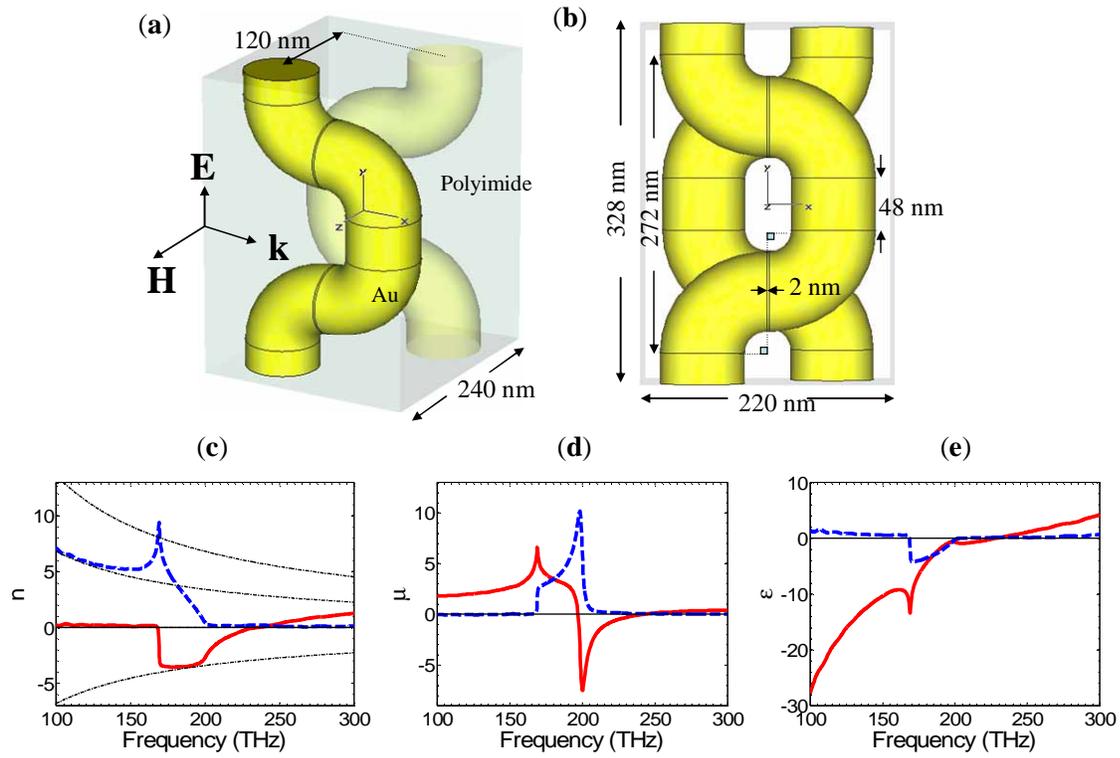

**Figure 1** (Color online) (**a-b**) Unit cell with parameters and the incident field configuration for the 1D NIM consisting of one pair of SRRs and (**c-e**) the corresponding retrieved effective parameters using HEM approximation. Solid red lines indicate the real parts and dashed blue lines indicate the imaginary parts. Magnetic permeability and electric permittivity are simultaneously negative in the region of negative index of refraction.

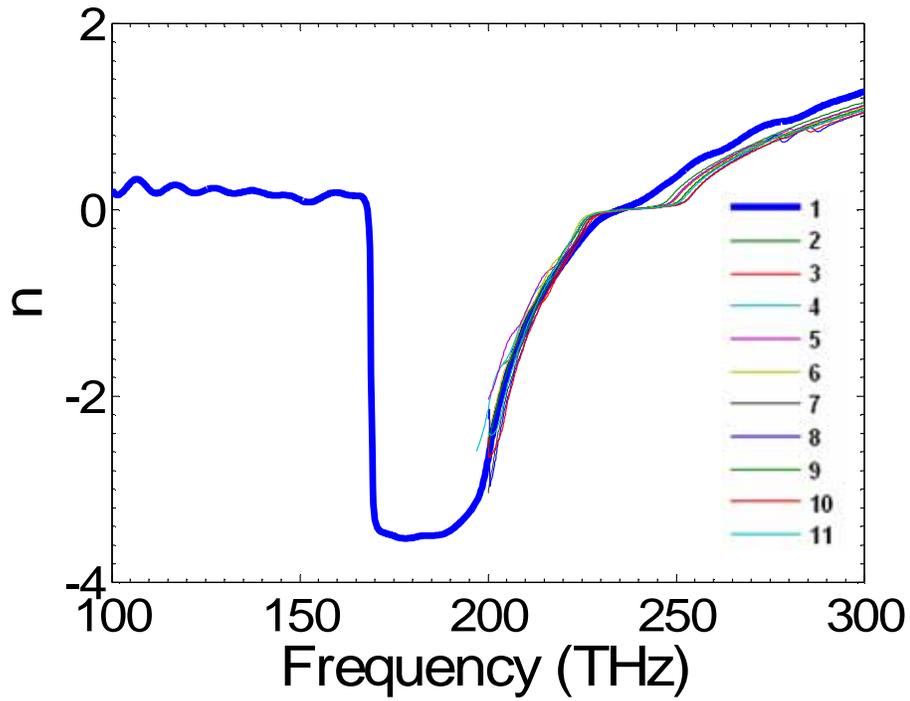

**Figure 2** (Color online) Retrieved effective index of refraction for different lengths of unit cells (1 to 11) plotted in the same graph. Curves for multiple unit cells (thin lines) tend to coincide except for the initial shift from the single unit cell curve (thick blue line) to the two unit cell curve.

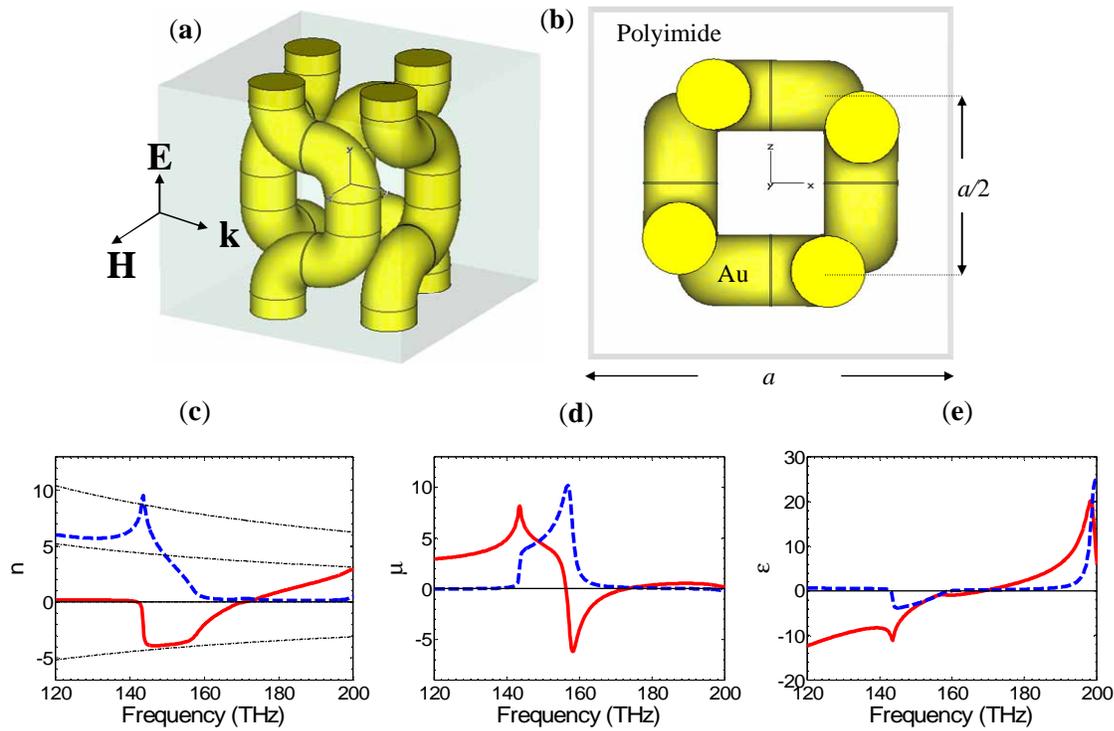

**Figure 3** (Color online) (**a-b**) Unit cell and the incident field configuration for the 2D NIM consisting of two pairs of SRRs and (**c-e**) the corresponding retrieved effective parameters using HEM approximation. Solid red lines indicate the real parts and dashed blue lines indicate the imaginary parts. Magnetic permeability and electric permittivity are simultaneously negative in the region of negative index of refraction.

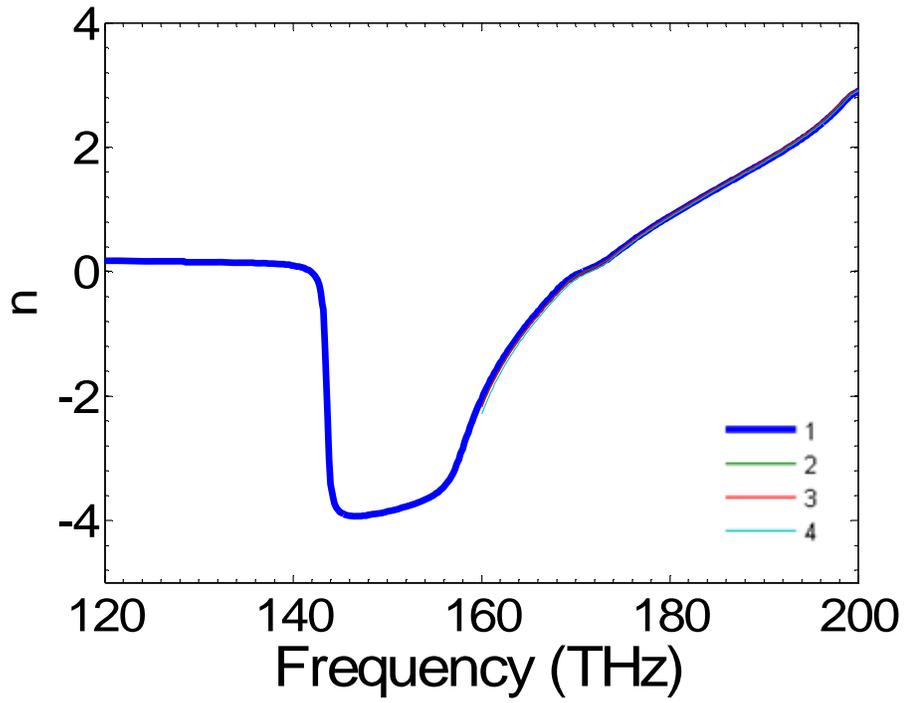

**Figure 4** (Color online) Retrieved effective index of refraction for the first four lengths (i.e., 1 to 4 unit cells) plotted in the same graph. Curves for multiple unit cells (thin lines) coincide even with that of single unit cell (thick blue line).